# Massively Parallelizable Approach for Evaluating Signalized Arterial Performance Using Probe-based data

*Subhadipto Poddar[a], Pranamesh Chakraborty[a], Anuj Sharma[a], Skylar Knickerbocker[a], Neal Hawkins[a]*

[a]Civil, Construction, and Environmental Engineering Department, Iowa State University, Ames, Iowa

A B S T R A C T

The effective performance of arterial corridors is essential to community safety and vitality. Managing this performance, considering the dynamic nature of demand requires updating traffic signal timings through various strategies. Agency resources for these activities are commonly scarce and are primarily by public complaints. This paper provides a data-driven prioritization approach for traffic signal re-timing on a corridor. In order to remove any dependence on available detection, probe-based data are used for assessing the performance measures. Probe-based data are derived from in-vehicle global positioning system observations, eliminating the need for installing on-field traffic infrastructure. The paper provides a workflow to measure and compare different segments on arterial corridors in terms of probe-based signal performance measures that capture different aspects of signal operations. The proposed method can serve as a tool to guide agencies looking to alter their signal control.

The methodology identifies a group of dynamic days followed by evaluation of travel rate based upon non-dynamic days. Dynamic days represent the variability of traffic on a segment. Performance measures on non-dynamic days include Median Travel Rate, Within-Day Variability of travel rate, between-days variability of travel rate Minimum Travel Rate Dispersion, and two variables which include Overall Travel Rate Variabilities. Consequently, a corridor having high number of dynamic segments along with poor performance during normal days would be a candidate for adaptive control. A case study was conducted on 11 corridors within Des Moines, Iowa where Merle Hay Road and University Avenue were identified suitable for adaptive control.

*Keywords:*
Traffic Signal Retiming
Arterial Corridor
Signal Performance Measure
Travel Time
Probe Data
Anomaly Detection

## 1. Introduction

Transportation agencies install traffic signals in order to optimize traffic flow, reduce crash frequency, and prioritize particular roadway user type or movement (Chandler et al., 2013).

Federal Highway Administration (FHWA) states that the operation and performance of the 300,000 signals in the United States (US) are addressed predominantly on the basis of citizen complaints (FHWA, 2017). Recognizing that a complaint-driven process is inefficient, many transportation agencies have sought objective methods for identifying and prioritizing corridors that require signal re-timing or the implementation of advanced signal control systems.

Two-thirds of all distance driven each year are on roadways controlled by traffic signals. Poor traffic signal timing is a major cause of traffic congestion and delay (Morales, 1995). Crashes are pervasive with signalized intersection operations. National Motor Vehicle Crash Causation Survey asserts that nearly 19% of the crashes occurred on signalized intersections which create points of conflict (Choi, 2010). Another study found that almost 67% of fatalities occur on arterial corridors (Brozen & Shockley, 2016). The National Traffic Signal Report Card (NTSRC) shows that improvements are being made as the overall score has increased from a letter grade of D- in 2005 to D+ in 2012, but a lot of issues still remain to be addressed (National Transportation Operations Coalition, 2012). The report card showed that the two lowest scoring categories were management which secured D in traffic monitoring and data collection which secured F. This paper identifies measures to mitigate these problems.

In order to maintain and improve the traffic signals, they are to be controlled in a sound and appropriate way to "move people through an intersection safely and efficiently" (Koonce et al., 2010). There are three primary operational modes for traffic signals: pre-timed control, semi-actuated control, and fully actuated control (Koonce et al., 2010). Improving these types of signals need periodic signal retiming (Curtis, 2017; Gordon, 2010). The signal retiming is invoked through the following ways (Gordon, 2010):

- Major changes in the land use pattern.
- Public requests
- Traffic conditions like oversaturation and spillback of queue.
- Detector or traffic camera video which suggests that there are changes in volume and congestion.

Retiming involves minimizing a delay-based objective function or maximizing arterial throughput through progression (Gordon, 2010). These assumptions do not always represent the real scenario in the field (Day, Brennan, et al., 2010) and are constrained by technical and institutional difficulties (Gordon, 2010). To circumvent these situations, the Adaptive Signal Control Technology (ASCT) came into existence (Curtis, 2017).

The ASCTs tend to maximize the capacity of the existing system based upon the information collected from the field to reduce cost of the system users and the operating agencies. They are useful in locations with variations in traffic rather than repetitions in traffic (Stevanovic, 2010). They have been reported to reduce travel times by 35–39% (Sims & Dobinson, 1980), reduce stops by 28–41% (Hicks & Carter, 1997), and also reduce crashes by 35% (Anžek, Kavran, & Badanjak, 2005). However, they involve a high initial cost for installation – both in field and at traffic management center (Stevanovic, 2010). The initial tune up and additional sensors tend to make the task tedious and calls for maintenance. The typical cost of implementing adaptive control ranges from $6,000 to $65,000 per intersection (Sprague, 2012; Zhao and Tian, 2012) which restricts a city-wide implementation. Even with higher costs, one-third of the ASCT has been found to be problematic in oversaturated traffic conditions (Stevanovic, 2010). A few cases have been reported where the ASCT has been removed from a corridor as it showed deterioration in performance due to non-ideal sensor performance, lack of trained staff, or inability of adaptive algorithms to respond to the site traffic conditions (Stevanovic, 2010). Although under an ideal scenario, a city-wide ASCT could provide a more effective solution. Only a few cities have implemented city-wide adaptive control due to current costs and sensor dependencies and is common to find ASCT on a few selected corridors. This paper provides the methodology to evaluate a corridor in terms of need for ASCT and how to prioritize the remaining signals for re-timing. It focuses on identifying locations that require automatic movement of split times to minimize public complaint.

When data are not accessible or require manual collection, significantly increases the cost of retiming (FHWA, 2017). Performance evaluation has been done using two different data source categories – infrastructure dependent data sources Bluetooth, cameras, loop detectors, or any kind of fixed/mounted sensors and non-infrastructure dependent data sources: probe-based data.

Infrastructure dependent methods can be divided into two categories - Automated Traffic Signal Performance Measures (ATSPMs) which require high-resolution data from the traffic signal controller and sensor data which involve mounted sensors like cameras and Bluetooth. The ability to extract ATSPMs is only available in the newer traffic signal controllers bought within the last few years. Table 1 presents the performance measures obtained using high-resolution traffic signal data. Apart from ATSPMs, travel time has been shown to be a consistent measure of corridor performance (J. Li, van Zuylen, & Wei, 2014). Several automated travel time determination methods have been developed, including anonymous address matching, cellular phone subscriber identity matching, and automatic license plate number matching (Singer et al., 2017, Venkatanarayana, 2017, Quayle et al., 2010, Day et al., 2012). Many of these are based on vehicle re-identification using paired sensors placed at upstream and downstream of traffic flow. These have proven reliable in daily service, but require installation and maintenance of roadside sensors and tend to represent biased sample of traffic stream (Chitturi et al., 2014, Shaw and Noyce, 2014). Other drawbacks of controller-based data include underestimate the traffic volume if the queue extends too far beyond the farthest-upstream loop (Smaglik et al., 2007b, Li et al., 2014) and the difference in detection rates at Bluetooth stations based upon a vehicle's position (Vo, 2011).

**Table 1**
**Methods to measure performance measures of arterial corridors and intersections.**

| Performance Measure | Methods used to measure |
| --- | --- |
| Delay | Stop bar and advanced detectors[*] (Sharma and Bullock 2008; Sharma et al. 2007). Video recording[*] (Sharma & Bullock, 2008). High-resolution event data[*] (Day & Bullock, 2010). |
| Number of stops | Video recording[*] (Fernandes et al., 2015). Connected Vehicle[+] (Argote-Cabañero, Christofa, & Skabardonis, 2015). |
| (Maximum) queue length | Stop bar and advanced detectors[*] (Sharma and Bullock 2008; Sharma et al. 2007). Video recording[*] (Sharma & Bullock, 2008). Stop bar and advanced detectors[*] (Sharma and Bullock 2008; Stop bar and probe data combined[#] (Comert, 2013). Probe data[+] (Comert & Cetin, 2009) |

| Arrival Type, Arrival rate on green, Degree of intersection saturation, Volume/capacity ratio, Level of progression, Split failure | Stop bar and setback detectors[*] (Smaglik, Bullock, & Sharma, 2007). High-resolution event data[*] (C M Day and Bullock 2010; Christopher M Day, Sturdevant, and Bullock 2010; Christopher M. Day et al. 2014) |
|---|---|
| **\* infrastructure dependent; + non-infrastructure dependent; # combined** ||

The use of non-infrastructure dependent GPS-based probe data overcomes some of these limitations since acquiring the data requires no roadside infrastructure, and the data aggregation and de-identification ensures that all the privacy concerns are appropriately addressed. Probe-based signal performance measures have been reported to have higher accuracy rate than the controller based data (Alhajri, 2014; Lattimer and Glotzbach, 2012; Li, 2013). They are also used for determining measures as noted in Table 1. They are used to investigate relationships between travel time and travel time reliability for arterials (Remias et al., 2013; Haghani et al., 2010; Hu et al., 2015).

While travel time provides an excellent proxy of traffic conditions within one segment, the measurement has inherent aggregate-level comparison problems (Day et al., 2014). Hence the travel rate is used for the analysis. Mathematically, travel rate is inverse of speed. Travel rate can be easily displayed using a cumulative distribution function (CDF) (Mathew, Krohn, Li, Day, & Bullock, 2017).

In a study, the performance of the arterial corridors was ranked based upon composite score derived from normalized travel rate (Day et al., 2015) along arterial corridors in Indiana. Neither did they remove any dynamic days when the travel rate was significantly different than the normal days, nor did they group the intersections based upon similar geometric properties (like traffic volume) which might have skewed the rankings.

The present work addresses the above shortcomings by grouping and then evaluating the performance of the segments under similar geometric properties. This paper presents a massively parallelizable technique that can process large scale data (hundreds of gigabytes) within a few minutes and compute performance of any number of segments that make up a corridor. The analysis starts with empirical CDF plots of travel rate which are used to evaluate the performance of arterial corridors using the following measures:

- Identifying dynamic days using CDF plots. This ensures that all dynamic days are removed. A large number of dynamic days mean that traffic is not repetitive and that ASCT could be a viable alternative. Where appropriate, dynamic days can be evaluated separately to identify the extent of the anomalies and potential interventions relevant to unusual situations.

- Evaluating the travel rate metric for normal days uses Median Travel Rate (MTR), Within-Day Variability (WDV) of travel rate, between-days variability of travel rate (Minimum Travel rate Dispersion (MTD), and two Overall Travel rate Variabilities (OTV_POLY and OTV_LINEAR)). The final prioritization threshold is based upon dividing the segments by geometric characteristics – annual average daily traffic per lane and intersection density. On these 5 parameters, a principal component analysis was conducted to identify segment performance as good or poor.

On these two measures, the need for adaptive control is determined. If a corridor is highly dynamic and performing poorly on normal days, indicates locations where adaptive control should be implemented first, followed by the ones which are performing poorly only on normal days.

The poor overall grade on NTSRC is a reflection of deficient signal retiming (National Transportation Operations Coalition, 2012). The measures used in this study will serve as a guideline to the different agencies who can evaluate the performance regularly, identify the problematic segments along corridors, and come up with quicker solutions, leading to better monitoring of the traffic signals which would the overall score in future evaluation.

The rest of the paper is broadly divided into three categories. First, it provides the details of the data used in this study and methodology adopted to identify performance of the segments. Then a case study is computed using this methodology to compare segments within the Des Moines, Iowa metro. Finally, a conclusion is drawn based upon the results observed for the case study.

## 2. Data and Methodology

### 2.1 Data Description and Data Reduction

This analysis was considered across 11 major arterial corridors within the Des Moines metropolitan area (refer Figure 1) consisting of 2nd Avenue, 22nd Street, Ashworth Road, Fleur Drive, Grand Avenue, Hickman Road, Jordan Creek Parkway, Merle Hay Road, Mills Civic Parkway, Southeast 14th Street, and University Avenue. The corridors were made up of 300 segments of which 51 were under existing adaptive control (22nd Street, Jordan Creek Parkway, Mills Civic Parkway, and University Avenue). Time period was from 5 am to 10 pm as at night there is low probe count and the adaptive control does not function during the overnight hours.

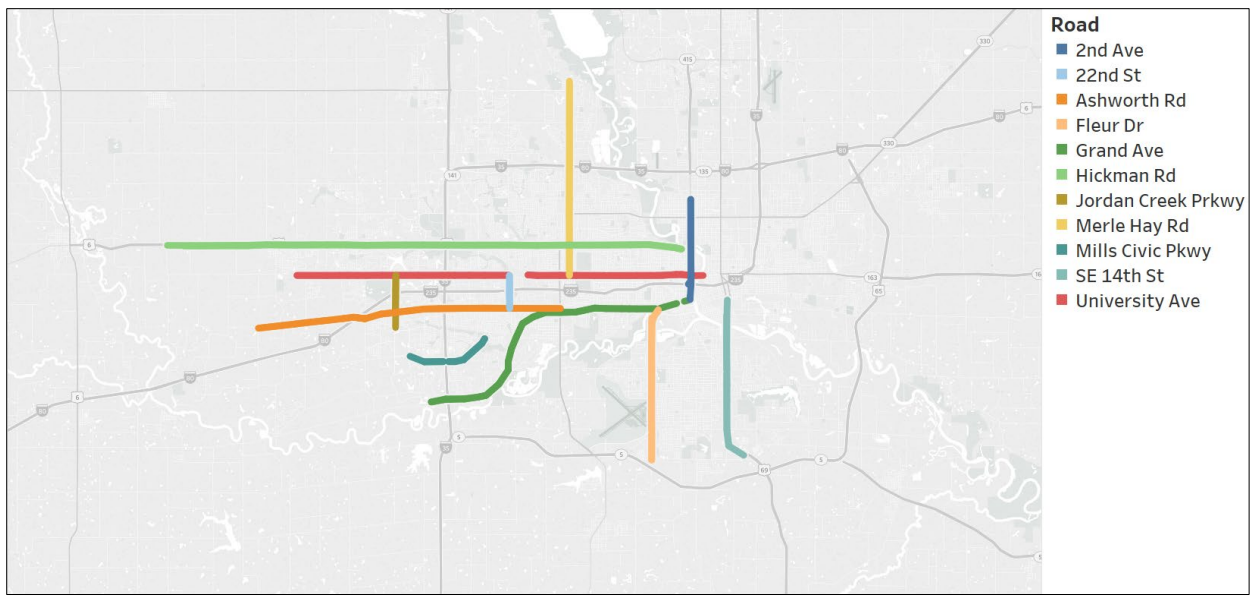

Figure 1. Location of the corridors in Des Moines.

Each corridor included multiple segments, defined by the probe data provider, ranging from 0.32 to 2.4 kilometer in length. The probe data for each segment included the average vehicular speed per direction per minute. The real-time data consists of that duration when there was presence of vehicle(s) on that road segment. These data were acquired from a commercial provider (INRIX) and the real-time data covered approximately 34% of the total duration.

Maintaining and analyzing the large dataset of 150 GB per month requires networked computers and parallel computational processing techniques. A distributed file system, Apache Hadoop was used for storing the information with the data sets stored in the form of smaller chunks to improve querying. Map-reduce codes were written in Apache Pig to extract the percentile values of speed for each segment. The parallelized computation ensured that filtering, mapping and reducing one month's worth of data, from the entire Iowa dataset, could be completed within 10 to 12 minutes.

## 2.2 Methodology

This section defines the methodology used as summarized in Figure 3. Raw data were acquired and converted to CDF plots for identifying the dynamic days on different segments. The dynamic days were used to identify dynamic corridors (arterials having a high percentage of dynamic day segments). After that, the remaining normal days were used to evaluate the performance measures in terms of MTR, WDV, OTV_POLY, OTV_LINEAR, and MTD. Using these, Principal Component Analysis (PCA) was carried out to characterize the behavior of each segment which helped to identify the corridors where the adaptive control is necessary. Based on this methodology, agencies can adopt their own thresholds by reproducing the methodology or they can directly use the thresholds obtained from the case study.

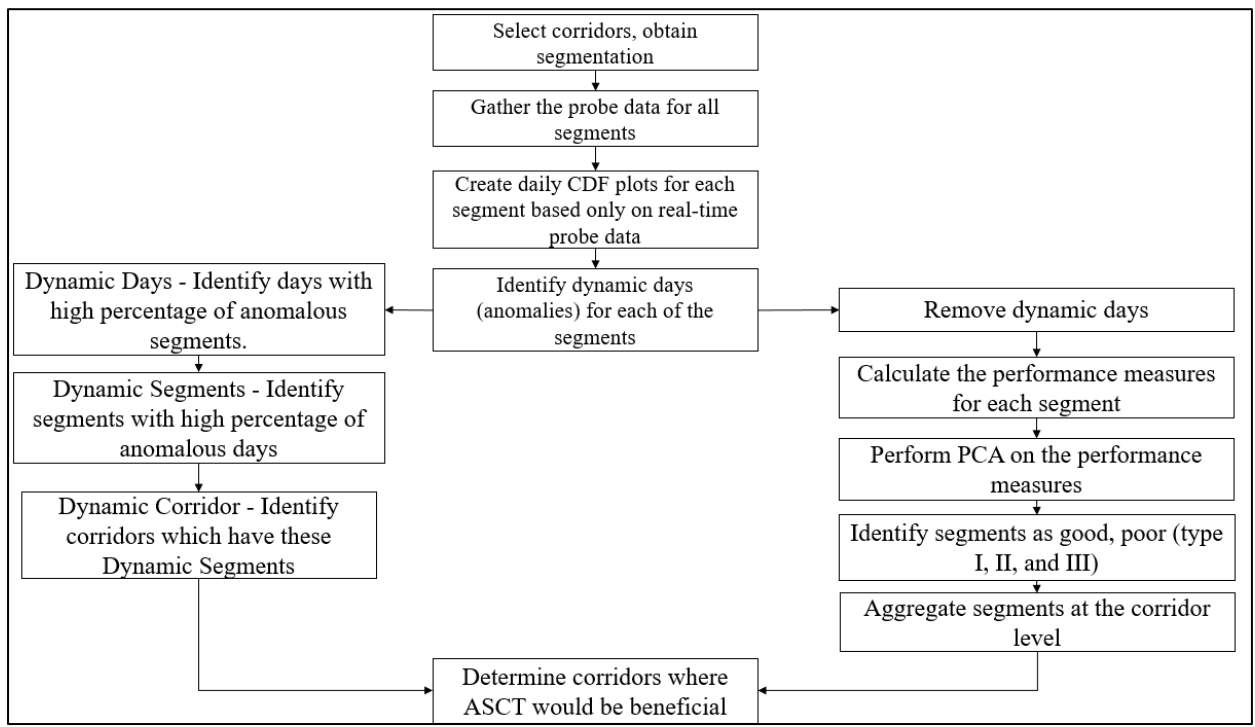

Figure 2: Flowchart of the methodology.

### 2.2.1 Identifying dynamic days using travel rate CDF plots

The variation of speed in a day can be represented as shown in **Figure 4**(a). For evaluating the daily performance, it was converted to a CDF plot. A CDF plots the probability of speed below a certain value. For example, if one says the 60th percentile is 34mph (refer to point 'a' in **Figure 4**(b)), implies that the probability of traveling below 60% of times the speed lower than 34 mph was observed at this segment.

Since they could not be added together to evaluate performance of entire corridor, speed CDFs were converted into travel time CDFs by dividing travel speeds with length of segments. This transformed point 'a' on **Figure 4**(b) into marked 'a' on **Figure 4**(c).

The travel time CDF cannot compare segments of unequal length. Thus, they were normalized by the length of the segment to obtain travel rate CDF plots. Fig 2(d) represents the travel rate CDF corresponding to the travel time CDF shown in Fig 2(c). The travel time CDF looks identical to the travel time CDF except for the values plotted on the X-axis. Further, point marked 'a' on Fig 2(c) got translated to an identical point in **Figure 4**(d).

Analysis was only conducted for those days that had at least 75 minutes of real-time probe data. In order to appropriate peak period data, days which had at least 30 minutes of data during each of the peak periods (6am-9am and 3pm-6pm during morning and evening respectively) were used for the analysis. For justifying these durations, segments lying on the Fleur Drive corridor were randomly chosen. Two filters were applied – variation of total minutes in a day (50, 75, and 100 minutes) and variation of the peak hour duration (30, 45, and 60 minutes). It was determined that there was no change in number of days for the total minutes of 50, 75, and 100 minutes. In the case of peak hour duration, the Kolmogorov–Smirnov test failed to show any significant difference between the three sets of data of 30, 45, and 60 minutes. This meant that the values chosen for the analysis were justified. This filter removed approximately 122 days per segment from the analysis.

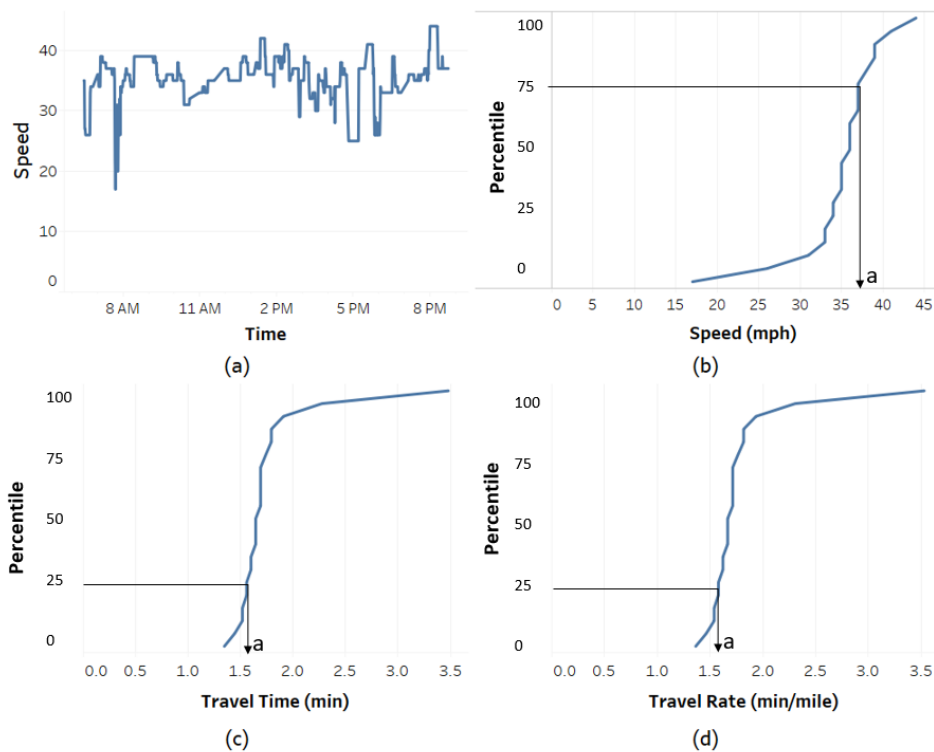

Figure 3. Plots of (a) Speed Variation, (b) Cumulative Distribution Plot for speed, (b) Cumulative Distribution Plot for travel time, and (c) Cumulative Distribution Plot for travel rate for a segment throughout a day.

Subsequently, the dynamic days were identified for each segment. The purpose of detecting dynamic days is to remove inaccurate travel rates from the analysis and to identify the necessity of a demand-driven signal re-timing. Dynamic days were abnormal days when the travel rate differed from normal due to events such as special events, construction, or adverse weather. The detection of dynamic days started with the daily travel rate CDF plots for each segment. Figure 5(a) displays an example plot of the travel rate CDFs for the northbound direction of the Water Works Park on the Fleur Drive corridor. Plots of different segments, on the same corridor and same direction, were further merged together to create a travel rate plot representative of the entire corridor. The principle of commonotonicity was applied to achieve the resultant plot. The principle states that for monotonically increasing functions, resultant $n^{th}$ percentile can be obtained as:

$$Res_n = \sum_{j=1}^{j=p} n(j) \tag{1}$$

where $Res_n$ represents the resultant $n^{th}$ percentile of the segment obtained by adding all the $n^{th}$ percentiles for p days.

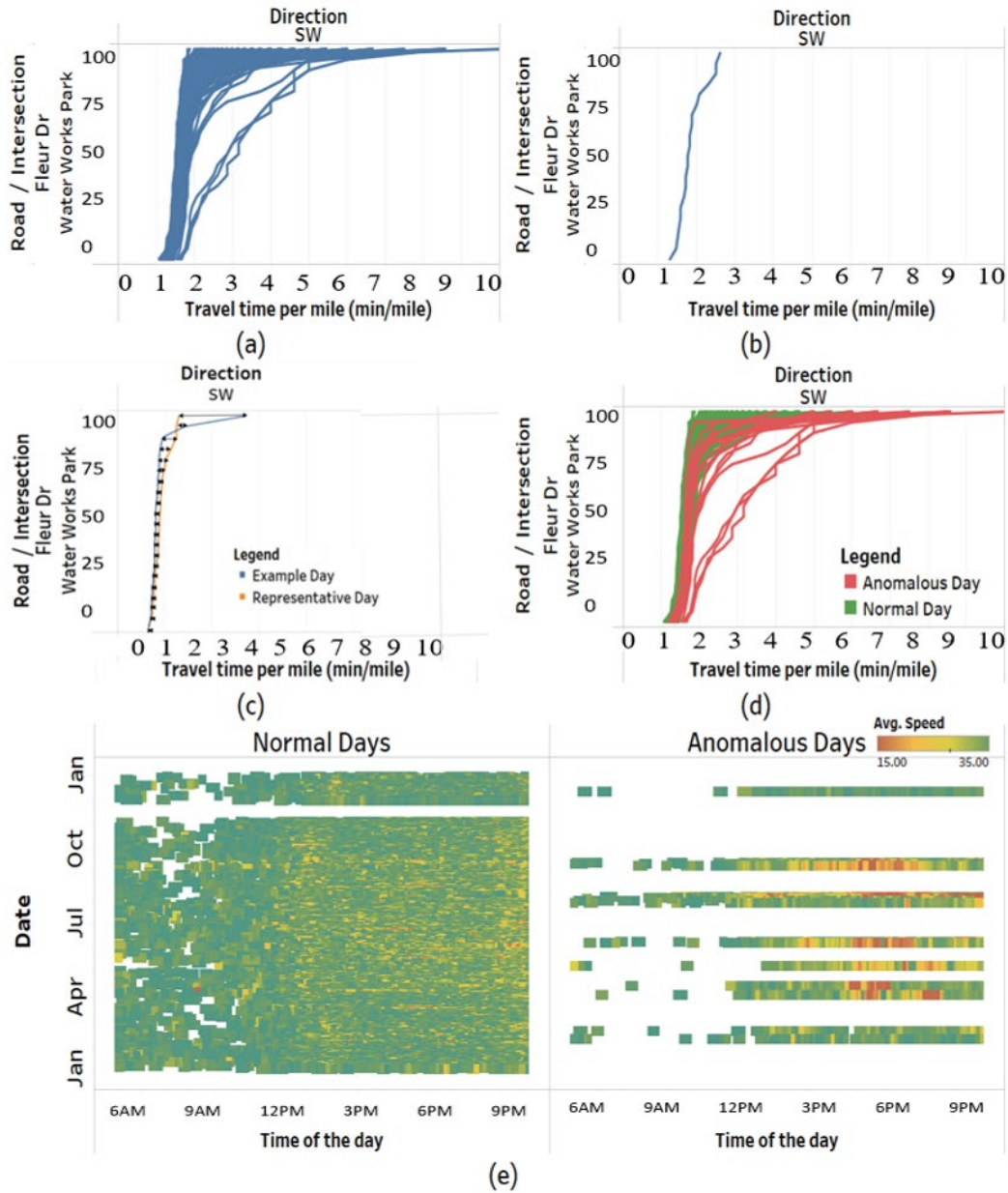

Figure 4. Plots showing (a) Daily CDF, (b) CDF plot of Representative day, (c) Difference between example day and representative day on CDF plots, (d) Normal and Dynamic days on CDF plots, and (e) Average 5-minute speed distribution for normal and dynamic days for Water Works Park, Fleur Drive.

Previous studies show that the above method has an error of 6% (Chen, Wang, & van Zuylen, 2010) for travel time. Applying this on the independent and identical plot for different days, for the same direction of a corridor, the representative plot for that direction of the corridor was obtained as shown in Figure 5(b) for the northbound direction of Fleur Drive. In order to compare the daily CDF plots to the representative CDF, the horizontal difference of the fifth percentiles (0, 5, 10, and so on) (Holland, 2007) between the representative day and each day were evaluated as shown for an example day on Figure 5(c). The mean and standard deviation of these differences were used as measures of dispersion. These measures were then utilized to evaluate the Local

Outlier Factor (LOF) score for each day. The details of the algorithm for Local Outlier Factor can be found in (Breunig, Kriegel, Ng, & Sander, 2000).

This was used to determine an elbow method cutoff which located the "higher than average" LOF values. Figure 5(d) shows the dynamic and normal days for a segment. A separate threshold was used in addition to elbow method to determine the cutoff value. A comparison of normal days and dynamic days can be visualized by the average 5-minute heat map plotted for the segment in Figure 5(e). Both Figure 5(d) and Figure 5(e) can be used to demonstrate that some of the dynamic days were mainly caused due to heavy congestion.

Based on the above algorithm, dynamic days were identified. A segment was defined as a dynamic segment when it had high volume of dynamic days with the value determined using the elbow-cutoff point and fixed LOF value. Dynamic segments were used to identify dynamic corridors which had one or multiple dynamic segments.

*2.2.2 Evaluating the travel rate metric for the normal days of the segment*

This step defines the performance measures obtained from the travel rate CDF plots for each segment. First, the normal days were derived for each segment from the previous analysis. Then the overall nature of the CDF plots were described based on their location (MTR), spread (WDV), and the overall shape (OTV_POLY, OTV_LINEAR, and MTD). These measures are further defined as:

- MTR – It represents the location measure of the CDF plots. It was the median time required to cross one mile of the segment. MTR was obtained as the 50th percentile of the median of each day's travel rate. The median travel rate is shown as point "a" in Figure 8 below.
- WDV – This represents the spread of the CDF plots. The WDV was the difference between the median 95th and 5th percentiles of a segment and reflects a segment's daily variability. High WDV values meant that there was significant fluctuation in travel rate for that segment. The WDV for a segment is shown as point "b" in Figure 8 below.

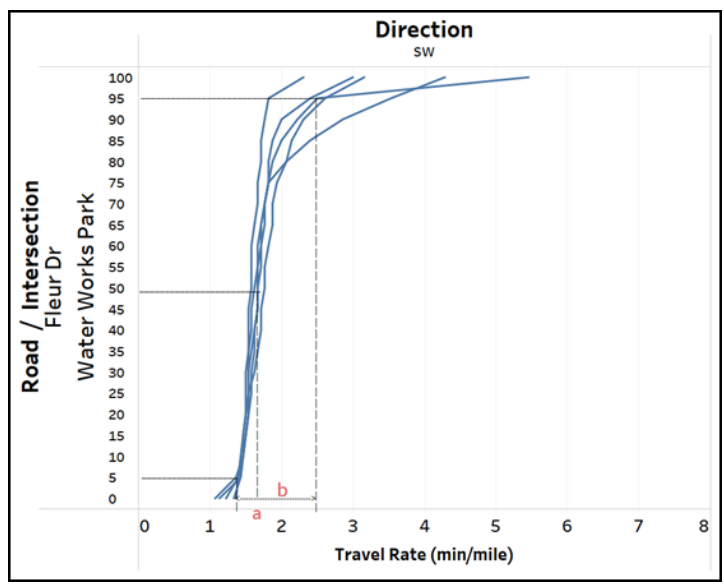

Figure 5. Plot showing the MTR as point "a" and WDV as point "b".

- OTV_LINEAR, OTV_POLY, and MTD – The variation in the shape of the CDF plots were captured using three measures. In order to obtain them, first, a 90% confidence envelope was obtained. This was done by joining the 5th and 95th percentiles of each 5th percentile (0, 5. 10, … 100) values of the travel rate CDF data as shown in Figure 9. Next, horizontal intermediate quantile differences were calculated at each 5% interval for twenty-one percentiles (0, 5, 10, and so on) (Holland, 2007) and a second-degree polynomial equation (quadratic) was fitted to the resulting data points as shown in Figure 9. Thus the points which are 'x', 'y' and 'z' change accordingly from Figure 9(b) to Figure 9(c). The equation is of the form:

$$Y = OTV\_POLY * x^2 + \text{OTV\_LINEAR} * x + MTD \tag{2}$$

Where, the coefficient OTV_POLY represented the quadratic nature of the change of quantiles, the coefficient OTV_LINEAR represented the linear change of quantiles, and MTD referred to the variability for the fitted travel rate at the 0th percentile.

OTV_POLY and OTV_LINEAR can have different combinations in the real case scenario. However, in the analysis, the former was always seen to be positive, the latter was seen to be negative, and the focus of the parabolic shape obtained was within 0 to 100. This gave rise to Figure 9(c). This meant that the difference of the percentile would first decrease for some percentile values and then keep on increasing. The coefficient MTD represented the variation of the travel rate during free flow conditions. High values of MTD meant that the free flow variation in travel time was significantly high.

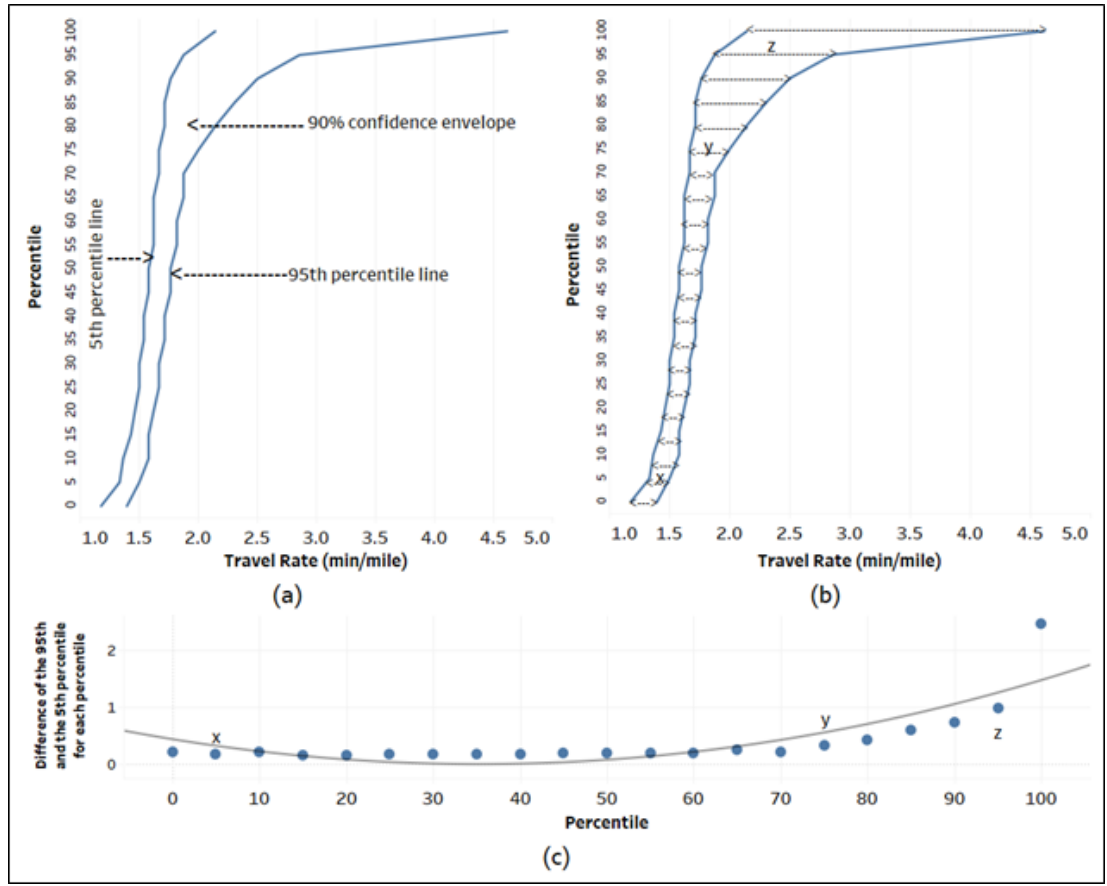

Figure 6. Figures showing (a) the 95% and 5% confidence lines and the 5% envelope, (b) The difference between the 95th percentile and the 5th percentile lines and c) The quadratic plot using the difference of the 95th and 5th percentile lines color-coded by the percentile values.

Using these five parameters (MTR, WDV, OTV_POLY, OTV_LINEAR, and MTD), the poor performing segments were identified. Since each segment had variability in traffic flows and number of intersections, they had to be divided based upon the following geometric parameters: Annual Average Daily Traffic per lane (AADT density) and number of intersections per mile of a segment (intersection density or ID). To divide them into a similar type of geometric performance, an agglomerative clustering algorithm was used (Pedregosa et al., 2011). Complete linkage which provided the maximum distance between two sets was used to cluster the elements as this would ensure that the groups were far apart from each other. In order to determine the appropriate number of intersection groups, the average silhouette score was determined for different cluster groups (Pedregosa et al., 2011). This is calculated based on the mean intra-cluster distance and the mean nearest-cluster distance for each segment.

For each of these groups, PCA was conducted using the five parameters of each segment. Based on the variance explained by the principal components (refer to Figure 10), three types of poor performing segments were identified.

- Poor Type I – These segments had all the measures performing poorly.
- Poor Type II – These segments had the WDV, OTV_POLY, OTV_LINEAR, and MTD parameters performing poorly.
- Poor Type III – These segments only had the MTR parameter performing poorly.

Segments can behave as good, poor type I, poor type II or poor type III. Based on them, the segments which are Poor Type – I would be the worst ones in performance.

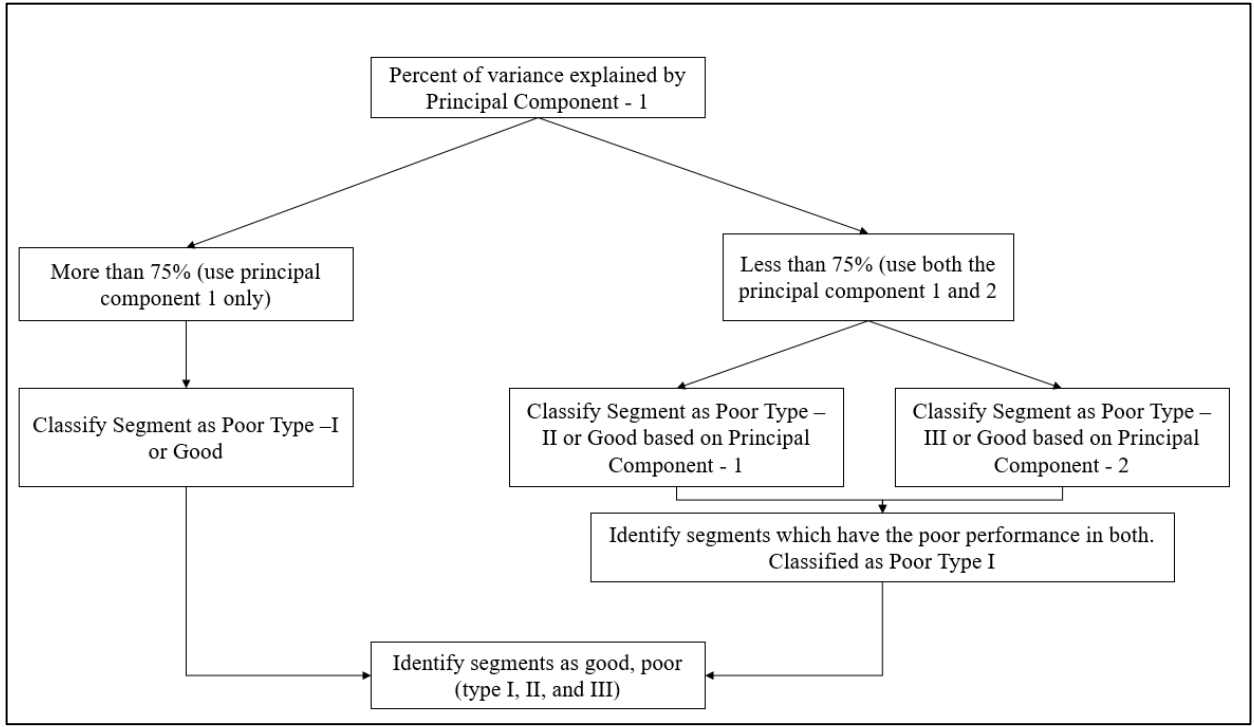

Figure 7. Performance of the segments based upon percentage of variance explained by the Principal Component(s).

Segments were finally aggregated in terms of the percentage of different behavior at a corridor level. Dynamic corridors and the performance of the corridors on normal days were used to judge the overall performance of any corridor. The corridors which were dynamic and poorly performing on normal days should be the first ones where the adaptive control may be implemented. This can be followed by the corridors which were performing poorly on normal days because a corridor is expected to perform well on normal days. This information can be used to support the traffic engineer's decision making when considering adaptive control as well as prioritizing the remaining signals that need retiming.

## 3. Results and Discussions

This section describes the application of this methodology to the Des Moines metropolitan corridors identified in Section 2.1.

### 3.1 Dynamic Days for Different Locations

As used in (Cheng, 1995), the days with a LOF value greater than 2, and also those beyond the cutoff point were seen to be significantly different from the others for most of the segments and was determined to be the threshold for dynamic days. Using this ideology, the dynamic days were identified for each segment, collected daily, and some of the top listed days are noted in Table 2. It was observed that weather was the major cause of dynamic behavior, suggesting that the speed of traffic was highly influenced by weather in the state of Iowa. The percentage of dynamic days show the variability in traffic demand. Higher percentage of dynamic days correspond to wider variability of travel rate over that segment. Analysis of the dynamic days revealed that the highest dynamic days were located on Grand Avenue and Ashworth Road. A segment is defined as a dynamic segment if it had more than 6% of dynamic days which was determined by the elbow cutoff point on the percentage of dynamic days. Dynamic segments were used to identify dynamic corridors as shown in Table 4, which identified Ashworth Road and Grand Avenue corridors as having the highest percentage of dynamic segments, which is apt as the former is influenced by the largest church in the state and later one by major entertainment events within the downtown area.

**Table 2: Reasons for some of the top dynamic days.**

| Rank | Date | Percentage of segments with dynamic days | Reasons |
|------|------|------------------------------------------|---------|
| 2 | 11/26/2016 | 17.78 | Thanksgiving weekend |
| 3 | 11/27/2016 | 17.39 | Thanksgiving weekend |
| 5 | 9/5/2016 | 15.38 | Labor Day |
| 6 | 12/17/2016 | 15.29 | Snow |
| 7 | 1/19/2016 | 14.68 | Snow |
| 8 | 1/24/2016 | 13.24 | Fog |
| 10 | 9/3/2016 | 12.40 | Des Moines Triathlon |
| 11 | 12/24/2016 | 12.33 | Christmas Eve |
| 13 | 2/10/2016 | 10.78 | Snow |
| 19 | 7/20/2016 | 9.13 | Flashfloods |
| 20 | 12/11/2016 | 9.09 | Sleet |

**Table 3: List of Dynamic Corridors**

| Corridor | Direction | Number of segments analyzed | Percentage of Dynamic Segments |
|----------|-----------|-----------------------------|-------------------------------|
| Ashworth Rd | N/E | 10 | 50 |
| Ashworth Rd | S/W | 13 | 62 |
| Fleur Drive | N/E | 10 | 10 |
| Grand Ave | N/E | 21 | 52 |
| Grand Ave | S/W | 23 | 35 |
| Hickman Rd | N/E | 30 | 3 |
| Hickman Rd | S/W | 31 | 6 |
| Merle Hay Rd | N/E | 13 | 15 |
| SE 14th St | S/W | 8 | 13 |
| University Ave* | N/E | 13 | 15 |
| University Ave* | S/W | 11 | 18 |
| University Ave | N/E | 15 | 7 |
| University Ave | S/W | 15 | 20 |
| *Adaptive corridor | | | |

## 3.2 Evaluation of Travel Rate Metric

The travel rate metric was determined for the remaining normal days. The five parameters – OTV_POLY, OTV_LINEAR, MTD, MTR, and WDV were calculated for each segment. The segments were then grouped based on AADT density and Intersection Density. In order to have an optimum number of clusters, the maximum number of clusters was limited to 15. The average silhouette measure yielded an optimum cluster number (maximum average silhouette measure) of 9 for the given segments. Out of these, 3 were outliers and the remaining analysis was conducted using the other 6 clusters which were named according to their values as shown in **Table 5**. Also, 90% confidence bounds for each group are shown in Table 5.

**Table 4: Groups for AADT per lane and ID for different type of segments with 90% confidence bounds.**

| Type of segment | AADT Density | ID |
|-----------------|--------------|-----|
| High AADT, Low ID | 9900-16275 | 0-2.03 |
| High AADT, Medium-low ID | 6913-15350 | 2.65-4.52 |
| High AADT, Medium-high ID | 8000-15920 | 5.34-7.88 |
| Low AADT, High ID | 3001-6550 | 7.72-9.70 |
| Low AADT, Low ID | 1296-7450 | 0-2.45 |
| Low AADT, Medium-low ID | 2062-7855 | 2.91-5.42 |

These measures were then used to define the performance of the segments on normal days. In order to capture the variation using fewer and more precise parameters, PCA was calculated for all the segments within each group. The percentage of variation explained by the first two components is shown in Table 6. Based on the percentage of variance, it was seen that the groups: Low AADT with Low ID, High AADT with Low ID, and Low AADT with High ID (referred to as Group 1) required one principal component to explain the performance of the segment. On the other hand groups with High AADT with Medium-high ID, Low AADT with Medium-low ID, and High AADT with Medium-low ID (referred to as Group 2) required two principal components. The different types of poor performance meant that the segment did not behave well in some respect of the travel rate as seen in Section 2.2.2. For this analysis, the poor performance was defined as follows.

- Poor Type I –This type included all the segments that have:
    o A positive principal component 1 for Group 1 (as shown in Figure 12).
    o A positive principal component 1 and positive principal component 2 for High AADT with Medium-low ID.
    o A positive principal component 1 and negative principal component 2 for all other members of Group 2 (refer to Figure 13).

- Poor Type II – This type included all the segments that have:
    o Poor performing Group 2 segments with respect to principal component 1 only (refer to Figure 13). These are represented by positive value on principal component 1 only.

- Poor Type III – This type included all the segments that have:
    o Poor performing Group 2 segments with respect to principal component 2 only (refer to Figure 14). These are represented by positive value on principal component 2 for High AADT with Medium-Low ID negative value on principal component 2 for other group members of Group 2 (refer to Figure 13).

**Table 5:  Percentage of variance explained by the first two principal components for each segment type.**

| Segment Type | Principal Component 1 | Principal Component 2 |
|---|---|---|
| High AADT, Low ID | 77.62 | 16.11 |
| High AADT, Medium-low ID | 65.47 | 27.82 |
| High AADT, Medium-high ID | 73.29 | 21.51 |
| Low AADT, High ID | 81.77 | 11.73 |
| Low AADT, Low ID | 83.85 | 10.44 |
| Low AADT, Medium-low ID | 65.68 | 27.62 |

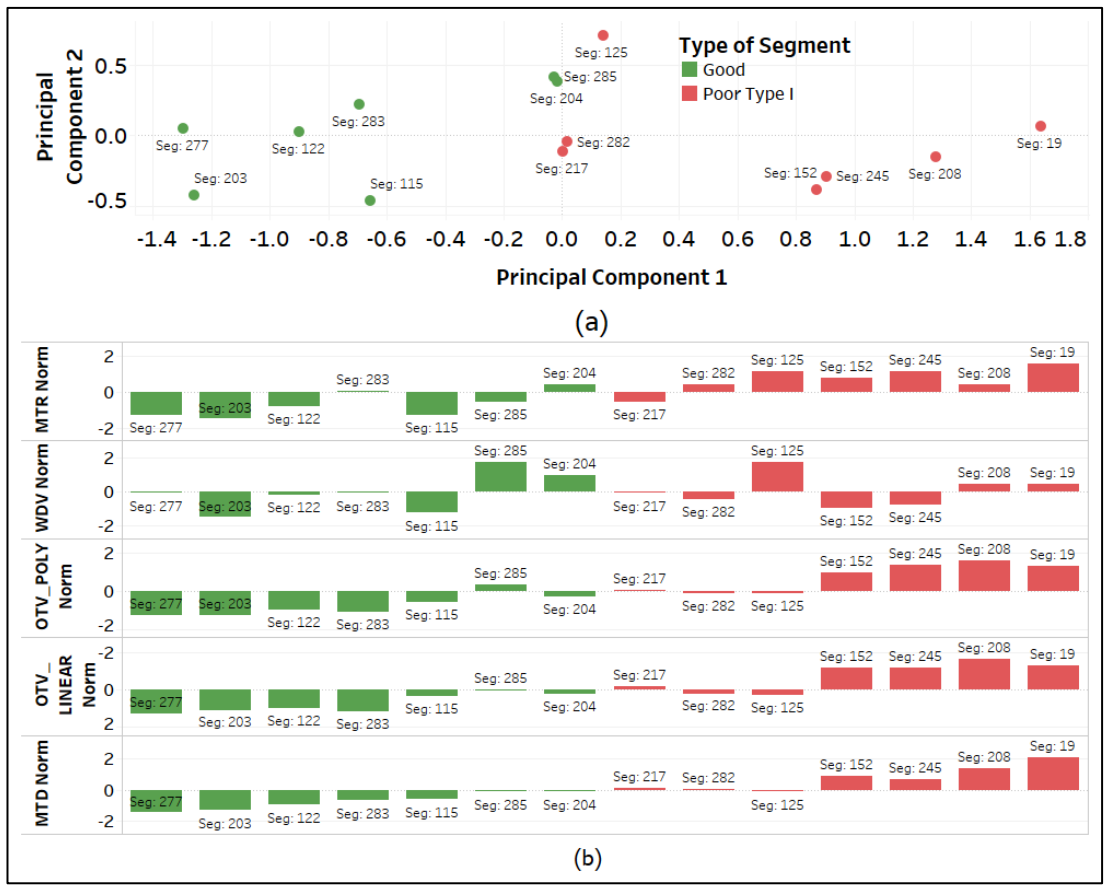

Figure 8. (a) Principal Component 2 versus Principal Component 1 for segments on Group 1 (Low AADT and High ID) colour coded based on PC 1and (b) Normalized values of MTR (MTR Norm), WDV (WDV Norm), OTV_POLY (OTV_POLY Norm), OTV_LINEAR (OTV_LINEAR Norm), and MTD (MTD Norm) for segments on Group 1 (Low AADT and High ID) colour coded based on PC 1.

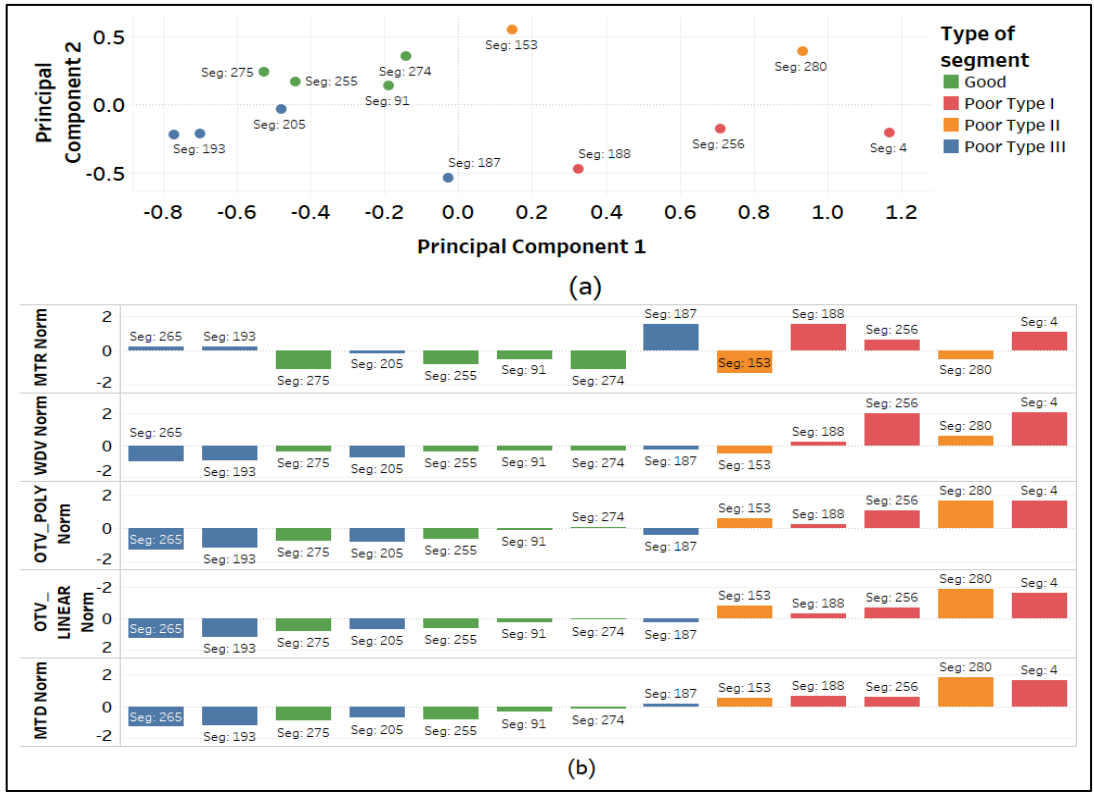

Figure 9. (a) Principal Component 2 versus Principal Component 1 for segments on Group 2 (High AADT and Medium-high ID) colour coded based on PC 1 and (b) Normalized values of MTR (MTR Norm), WDV (WDV Norm), OTV_POLY (OTV_POLY Norm), OTV_LINEAR (OTV_LINEAR Norm), and MTD (MTD Norm) for segments on Group 2 (High AADT and Medium-high ID) colour coded based on PC 1.

After classifying the segments for each category, they were accumulated at a corridor level as seen in Figure 14. Based on their current signal deployments, they were also grouped into adaptive and non-adaptive group to determine the performance of these groups separately. It was found that 22nd Street and Jordan Creek Parkway were the ones which remained problematic despite having ASCT installed on them. University Avenue, Merle Hay Road, and Grand Avenue were the non-ASCT based corridors that had high percentage of problematic segments. The main cause for 22nd Street was that the segments were very short in length (average of 0.27 mile) as compared to other corridors (average of 0.49 mile) which led to high variation in traffic. Jordan Creek Parkway, despite currently being under adaptive control, performed rather poorly as it is the major arterial corridor connecting University Avenue and Ashworth Road to Interstate 80. Like Ashworth Road, Jordan Creek Parkway serves both businesses, churches, and a regional mall which increases corridor travel times.

Finally in corridors were identified to implement ASCT on the existing non-ASCT corridors. Grand Avenue, University Avenue and, Merle Hay would be the three most suitable locations to be considered. However, it would be difficult to implement adaptive control on Grand Avenue as it is in the downtown area where it is expected to run on fixed time coordination patterns to serve all movements including pedestrian mobility. Hence, adaptive control would benefit University Avenue and Merle Hay Road. Also, a thorough investigation can be conducted on the Jordan Creek Parkway to determine why it is unable to perform adeptly under ASCT.

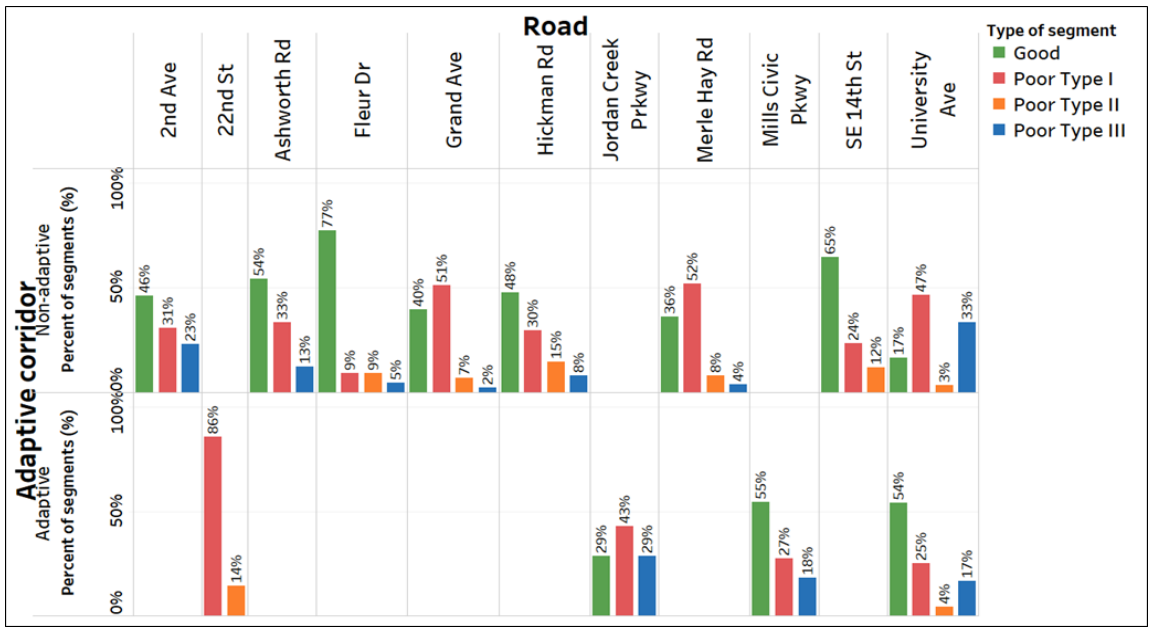

Figure 10. Type of segments for each corridor based upon their performance.

## 4. Conclusion

Probe-based real-time data are available from a number of vendors. This study uses such data source to determine segment performance and identify problematic segments on different arterial corridors. Using parallel computing, this method aims to compute these results from a large volume of data in a very shorter time.

Travel rate, defined as the travel time per mile, was determined as the parameter used to evaluate these measures. CDF plots of travel rate were constructed to represent the variability and behavior of the segments for each day. Unlike previous studies which used all days to capture the performance road segment, this study identified and removed the dynamic days for each segment of a roadway. Dynamic days were then evaluated separately to identify the extent of the anomalies and potential interventions relevant to unusual situations, such as special events, severe traffic incidents, extreme weather, and construction. After removing dynamic days, segments were classified into homogenous geometrical characteristics – Annual Average Daily Traffic per lane and number of intersections per segment. Five other measures were further extracted for these segments to quantify the overall performance of a segment - MTR, WDV, MTD, OTV_POLY, and OTV_LINEAR. PCA was further applied to these variables to reduce dimensionality and further classify the segments as good or poor.

A case study was conducted for eleven arterial corridors within the Des Moines metropolitan area in Iowa to identify their performance. The dynamic days were first extracted and the segments with the highest dynamic days were found on Grand Avenue and Ashworth Road. After removing the dynamic days, the remaining segments were analyzed for their typical behavior. Most of the problematic segments occurred on corridors which already have adaptive control such as 22nd Street and Jordan Creek Parkway. The most problematic segments, where adaptive control does not currently exist, were on Grand Avenue, Merle Hay Road, and University Avenue. As intersections near downtown tend to have fixed time coordination, the next two corridors suitable for adaptive are Merle Hay Road and the University Avenue.

The demonstrated methodology can be applied to identify problematic segments in the future. The tool can assist in identifying locations where delay is high, day-to-day traffic patterns are dynamic, or the minute-to-minute demands at signalized intersections are highly variable. Using this methodology, agencies can come up with their own thresholds or they can directly use these thresholds to determine the performance of the different segments. Future work can include applying this methodology to corridors, in other cities, to test these threshold values and findings. Through the measures defined, transportation agencies can easily automate the process of monitoring the performance of arterials so they can identify, screen, and prioritize signal retiming or traffic control modifications. These types of tools can support agency decision making, planning, and operational investments as they try to provide both throughput and safety for roadway users.

## Acknowledgment

The authors would like to thank the Midwest Transportation Center, the U.S. Department of Transportation (DOT) Office of the Assistant Secretary for Research and Technology, and Iowa DOT for sponsoring this research. The authors would also like to acknowledge the Federal Highway Administration (FHWA) that contributed state planning and research funds to the Iowa DOT as part of their funding and provided match funds for this project.

**Tables and Figures**

Table 1. Methods to measure performance measures of arterial corridors and intersections.

| Performance Measure | Methods used to measure |
| --- | --- |
| Delay | Stop bar and advanced detectors[*] (Sharma and Bullock 2008). Video recording[*] (Sharma & Bullock, 2008). High-resolution event data[*] (Day & Bullock, 2010). |
| Number of stops | Video recording[*] (Fernandes et al., 2015). Connected Vehicle[+] (Argote-Cabañero, Christofa, & Skabardonis, 2015). |
| (Maximum) queue length | Stop bar and advanced detectors[*] (Sharma and Bullock 2008). Video recording[*] (Sharma & Bullock, 2008). Stop bar and advanced detectors[*] (Sharma and Bullock 2008). Stop bar and probe data combined[#] (Comert, 2013). Probe data[+] (Comert & Cetin, 2009) |
| Arrival Type, Arrival rate on green, Degree of intersection saturation, Volume/capacity ratio, Level of progression, Split failure | Stop bar and setback detectors[*] (Smaglik, Bullock, & Sharma, 2007). High-resolution event data[*] (C M Day and Bullock 2010) |
| * infrastructure dependent; + non-infrastructure dependent; # combined | |

Table 2. Reasons for some of the top dynamic days.

| Rank | Date | Percentage of segments with dynamic days | Reasons |
| --- | --- | --- | --- |
| 2 | 11/26/2016 | 17.78 | Thanksgiving weekend |
| 3 | 11/27/2016 | 17.39 | Thanksgiving weekend |
| 5 | 9/5/2016 | 15.38 | Labor Day |
| 6 | 12/17/2016 | 15.29 | Snow |
| 7 | 1/19/2016 | 14.68 | Snow |
| 8 | 1/24/2016 | 13.24 | Fog |

| 10 | 9/3/2016 | 12.40 | Des Moines Triathlon |
|----|----------|-------|----------------------|

Table 3: List of Dynamic Corridors.

| Corridor | Direction | Number of segments analyzed | Percentage of Dynamic Segments |
|----------|-----------|------------------------------|--------------------------------|
| Ashworth Rd | N/E | 10 | 50 |
| Ashworth Rd | S/W | 13 | 62 |
| Fleur Drive | N/E | 10 | 10 |
| Grand Ave | N/E | 21 | 52 |
| Grand Ave | S/W | 23 | 35 |
| Hickman Rd | N/E | 30 | 3 |
| Hickman Rd | S/W | 31 | 6 |
| Merle Hay Rd | N/E | 13 | 15 |
| SE 14th St | S/W | 8 | 13 |
| University Ave* | N/E | 13 | 15 |
| University Ave* | S/W | 11 | 18 |
| University Ave | N/E | 15 | 7 |
| University Ave | S/W | 15 | 20 |
| *Adaptive corridor | | | |

Table 4: Groups for AADT per lane and ID for different type of segments with 90% confidence bounds.

| Type of segment | AADT Density | ID |
|-----------------|--------------|-----|
| High AADT, Low ID | 9900-16275 | 0-2.03 |
| High AADT, Medium-low ID | 6913-15350 | 2.65-4.52 |
| High AADT, Medium-high ID | 8000-15920 | 5.34-7.88 |
| Low AADT, High ID | 3001-6550 | 7.72-9.70 |
| Low AADT, Low ID | 1296-7450 | 0-2.45 |
| Low AADT, Medium-low ID | 2062-7855 | 2.91-5.42 |

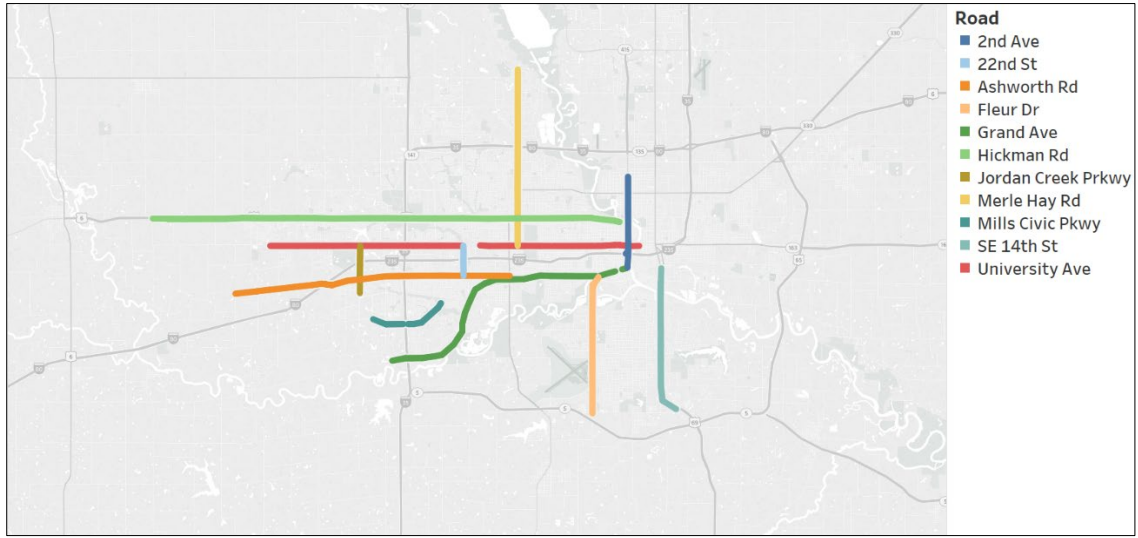

Figure 1. Location of the corridors in Des Moines.

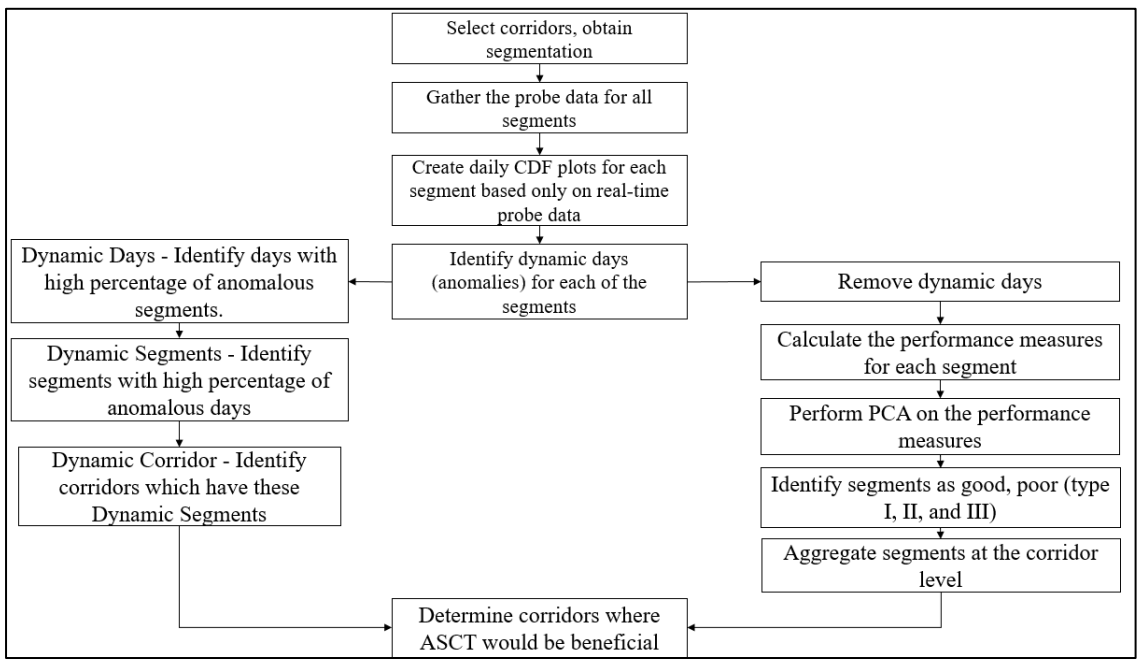

Figure 2. Flowchart of the methodology.

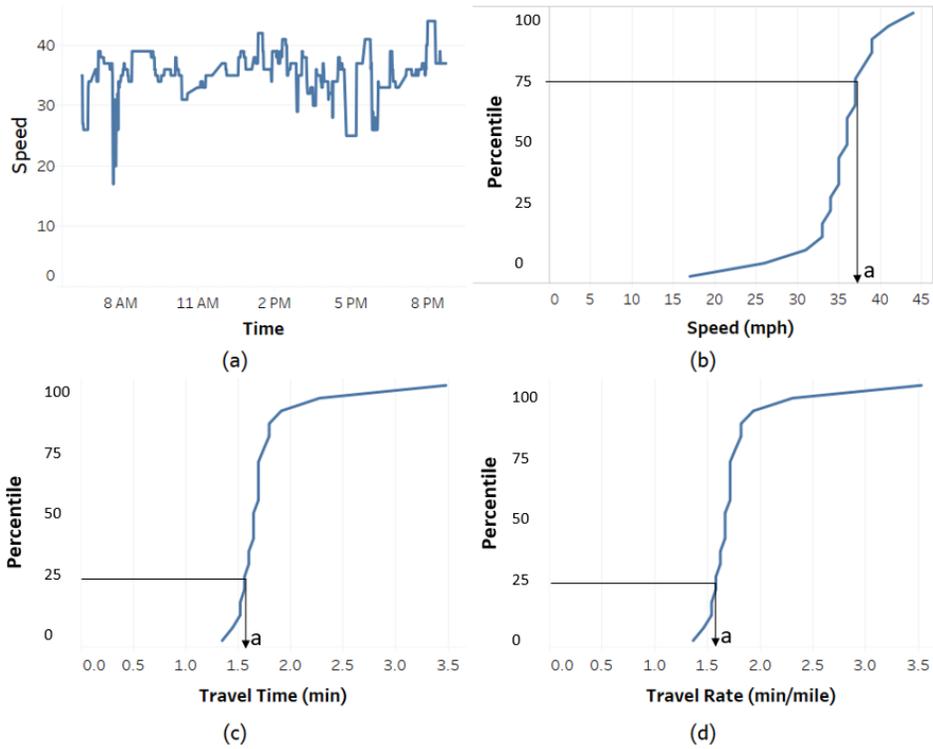

Figure 3. Plots of (a) Speed Variation, (b) Cumulative Distribution Plot for speed, (b) Cumulative Distribution Plot for travel time, and (c) Cumulative Distribution Plot for travel rate for a segment throughout a day.

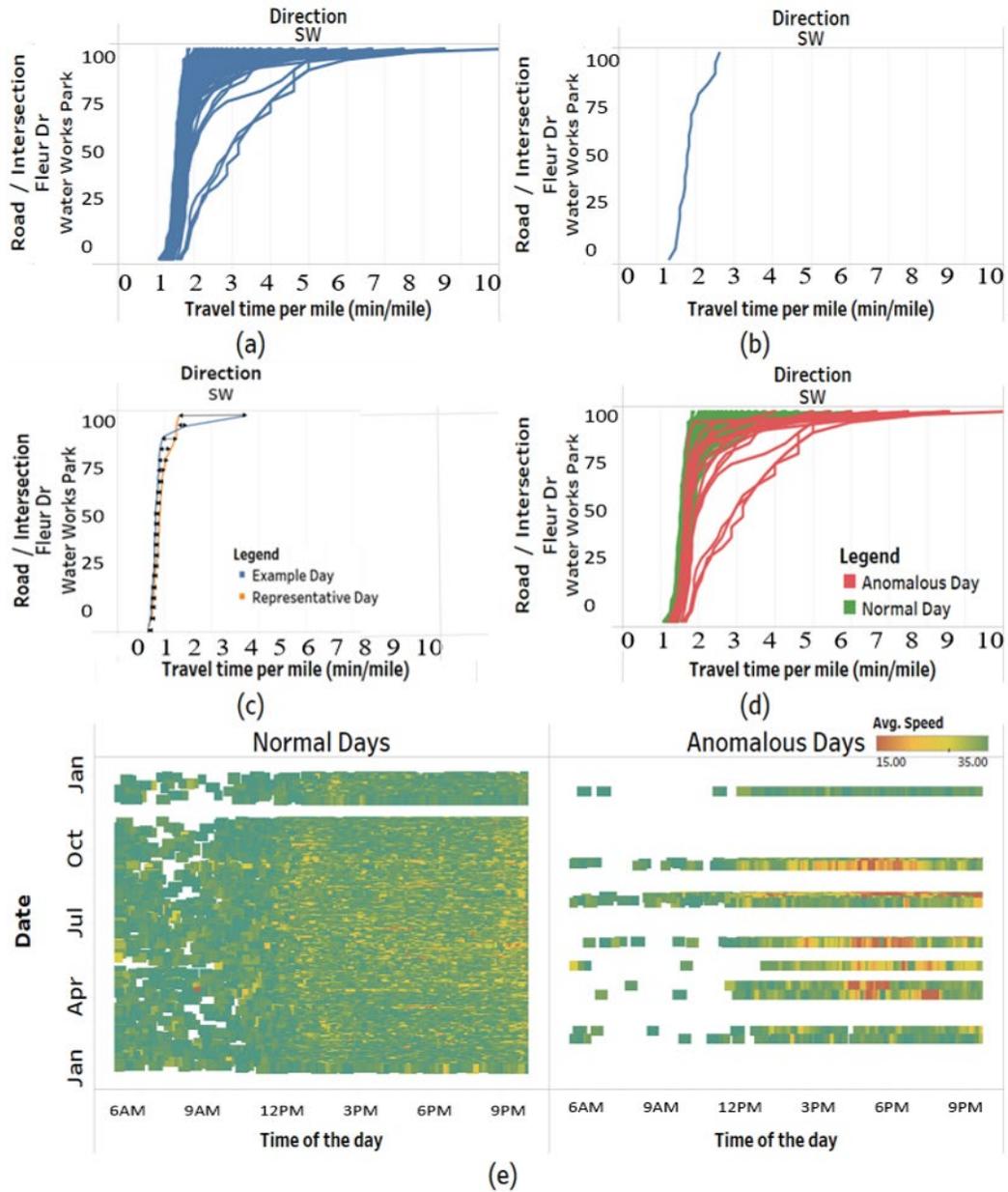

Figure 4. Plots showing (a) Daily CDF, (b) CDF plot of Representative day, (c) Difference between example day and representative day on CDF plots, (d) Normal and Dynamic days on CDF plots, and (e) Average 5-minute speed distribution for normal and dynamic days for Water Works Park, Fleur Drive.

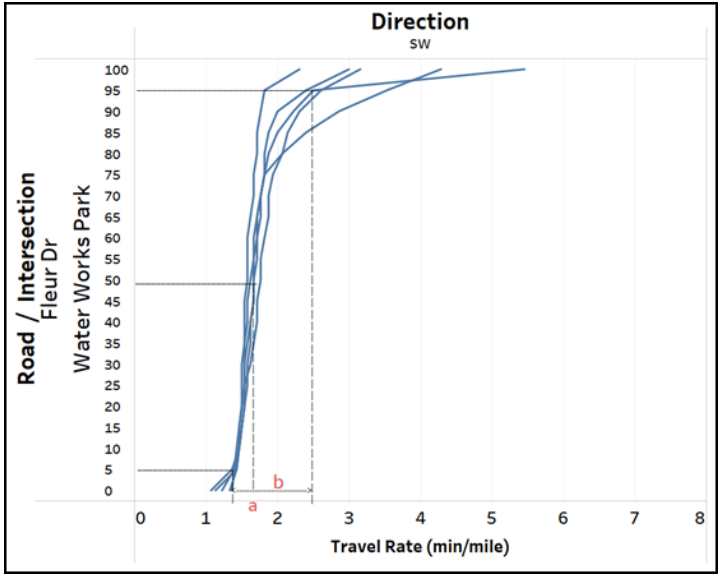

Figure 5. Plot showing the MTR as point "a" and WDV as point "b".

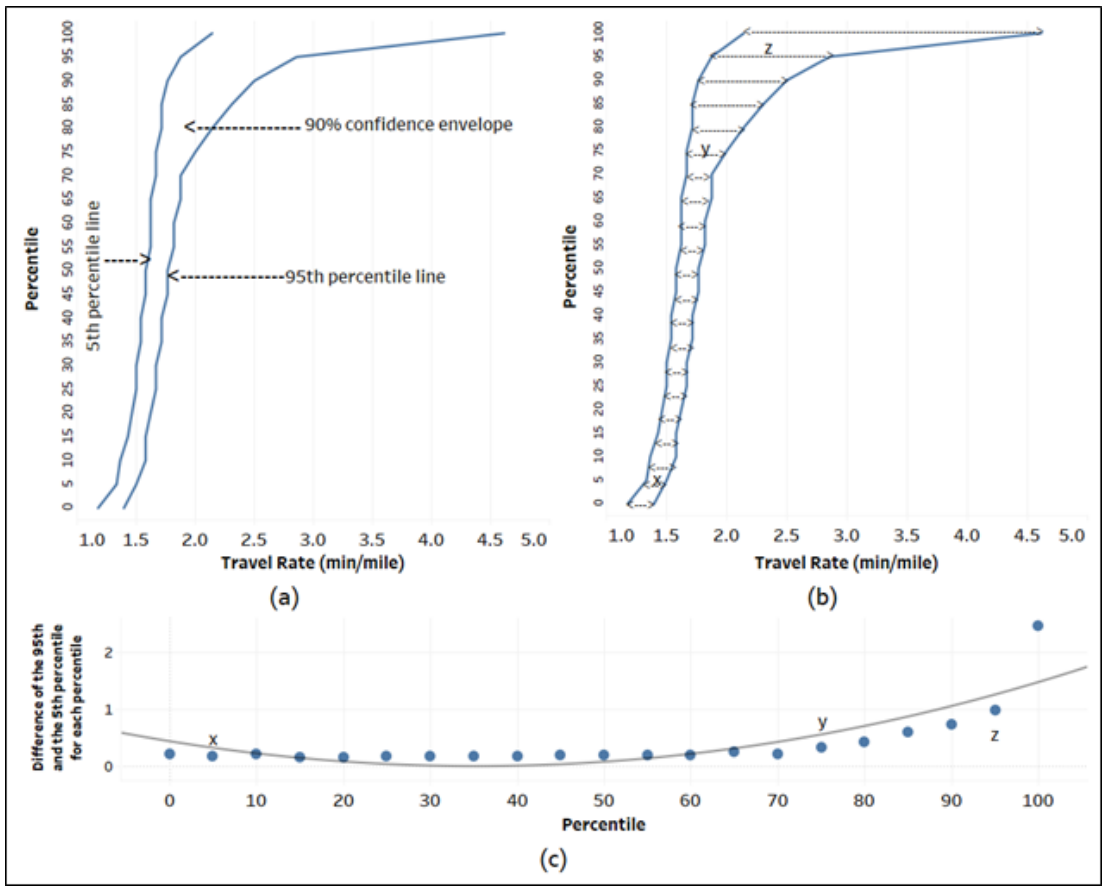

Figure 11. Figures showing (a) the 95% and 5% confidence lines and the 5% envelope, (b) The difference between the 95th percentile and the 5th percentile lines and c) The quadratic plot using the difference of the 95th and 5th percentile lines color-coded by the percentile values.

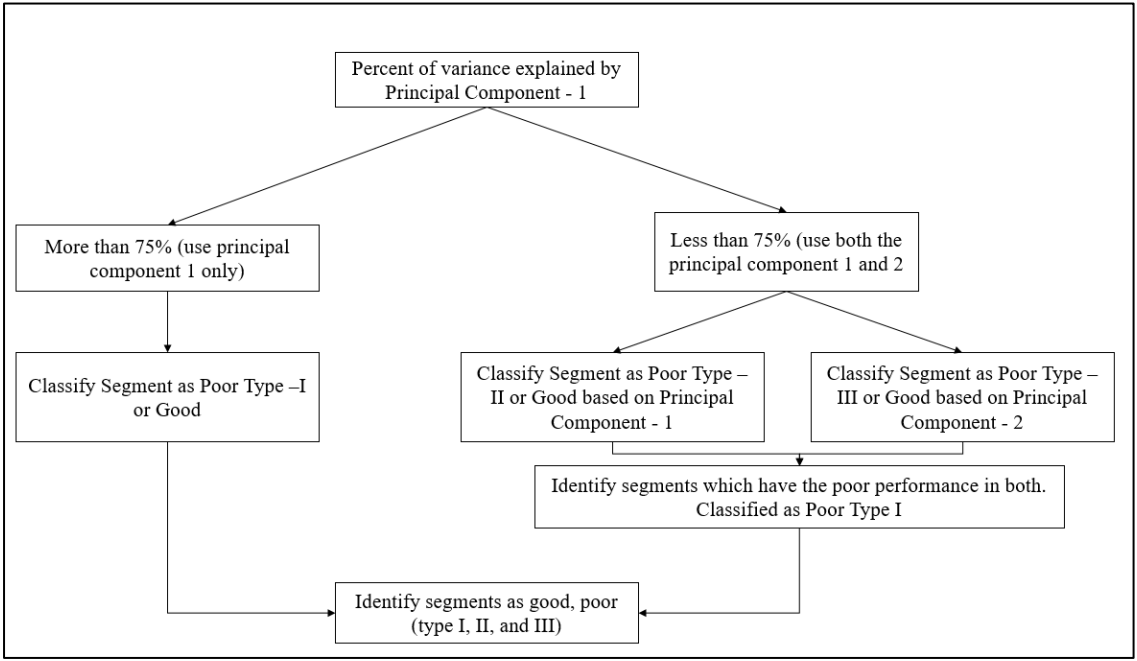

Figure 7. Performance of the segments based upon percentage of variance explained by the Principal Component(s).

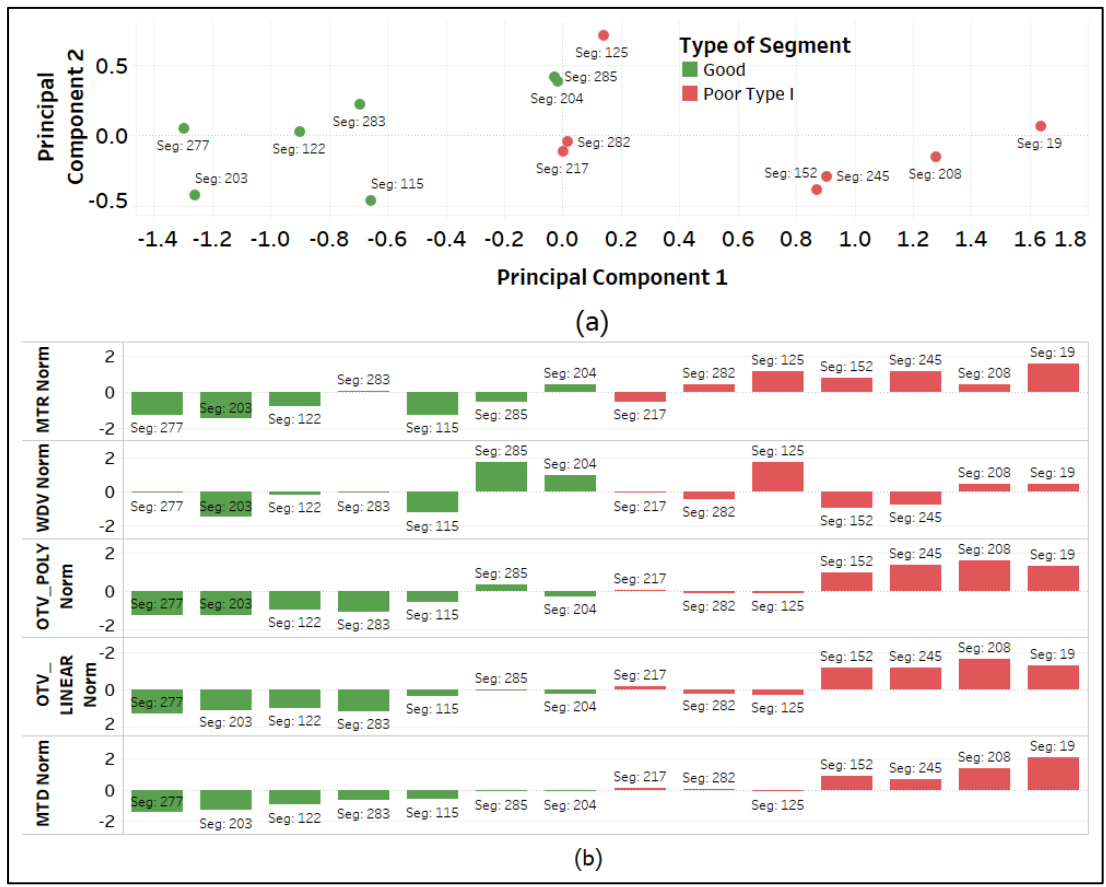

Figure 8. (a) Principal Component 2 versus Principal Component 1 for segments on Group 1 (Low AADT and High ID) colour coded based on PC 1and (b) Normalized values of MTR (MTR Norm), WDV (WDV Norm), OTV_POLY (OTV_POLY Norm), OTV_LINEAR (OTV_LINEAR Norm), and MTD (MTD Norm) for segments on Group 1 (Low AADT and High ID) colour coded based on PC 1.

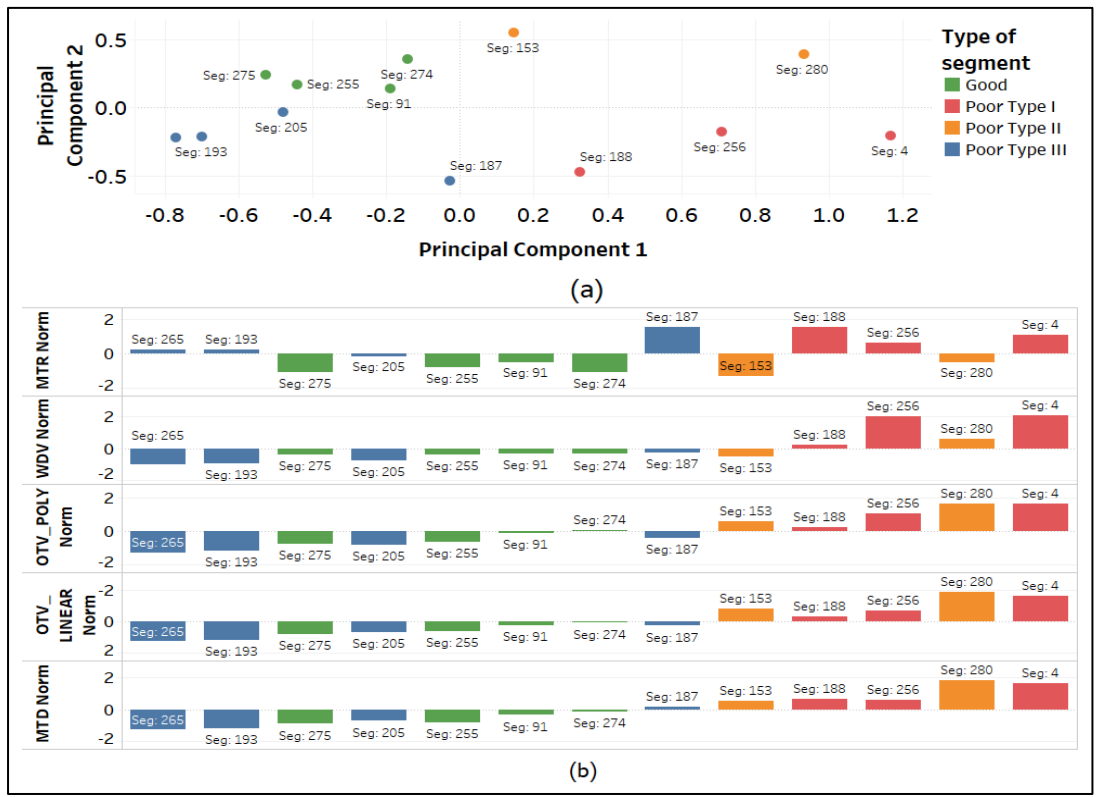

Figure 9. (a) Principal Component 2 versus Principal Component 1 for segments on Group 2 (High AADT and Medium-high ID) colour coded based on PC 1 and (b) Normalized values of MTR (MTR Norm), WDV (WDV Norm), OTV_POLY (OTV_POLY Norm), OTV_LINEAR (OTV_LINEAR Norm), and MTD (MTD Norm) for segments on Group 2 (High AADT and Medium-high ID) colour coded based on PC 1.

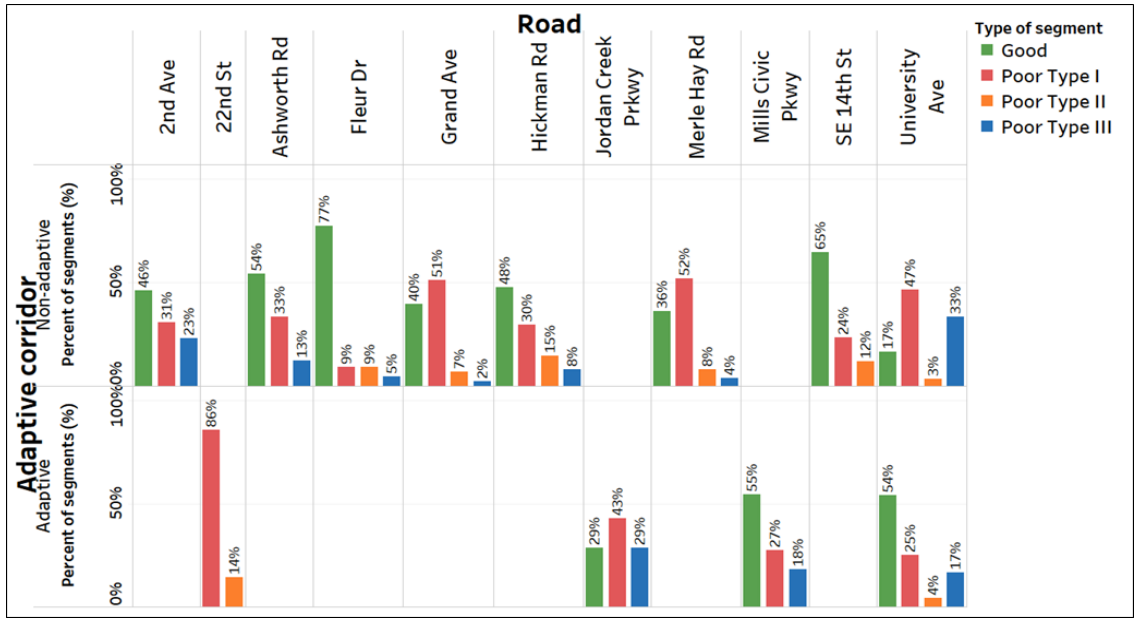

Figure 10. Type of segments for each corridor based upon their performance.